# METAHEURISTIC OPTIMIZATION FRAMEWORK FOR DRAG REDUCTION USING BIOINSPIRED SURFACE RIBLETS

**Aakash Ghosh**
Delhi Technological University, Delhi, India

**Priyam Gupta**
Delhi Technological University, Delhi, India

**Jayant**
Delhi Technological University, Delhi, India

**Raj Kumar Singh**
Delhi Technological University, Delhi, India

## ABSTRACT

*Emulating natural mechanisms in technology has become a very efficient technique to optimize and improve the current machinery. Riblets are such kinds of bio-inspired surface patterns which are seen on Sharks. Their geometric properties induce secondary flows resulting in lower drag, allowing Sharks to achieve such high speeds. Secondary flows are broadly subcategorized into Prandtl's first and second kind; the secondary flows induced by riblets belong primarily to Prandtl's second kind. These secondary flows have been proven to delay the boundary layer separation by reducing momentum losses. This paper aims to optimize the use of these bio-inspired patterns on an airfoil. By applying the ribletted surface on an airfoil, the boundary layer separation can be delayed, thus leading to a reduction in drag and increasing the critical angle of attack. Specific geometric properties such as the height and the wavelength of the riblets are varied to find the most optimal design. The design framework couples the optimization algorithm, computational fluid dynamics, and post-processing analysis. An Invasive Weed Model is implemented for optimization due to its commendable performance in converging close to global optima. The optimization process is initialized by generating a population of riblet profiles with randomly distributed geometry parameters. These profiles are then evolved over generations with the objective to minimize the drag, which is computed through Computational Fluid Dynamics simulations conducted on OpenFOAM. The optimized geometries show a significant reduction in drag as compared to bare airfoils. Lastly, a flow field analysis of the optimized geometry established using the genetic algorithm is done to understand the riblets-induced enhancement in the aerodynamic efficiency of the airfoil.*

## NOMENCLATURE

Place nomenclature section, if needed, here. Nomenclature should be given in a column, like this:

| | |
|---|---|
| α | Angle of Attack |
| h | Height of Riblet (mm) |
| s | Wavelength of Riblet (mm) |
| c | Chord of Airfoil (mm) |
| U | Flow Velocity (m/s) |

## 1. INTRODUCTION

Evolution is one of the most awe-inspiring processes in this universe. Evolution is a series of biological optimizations allowing several different biological beings to adapt and survive the forever changing atmosphere of Earth. One of these adaptations have been noticed on Sharks. Sharks are one of the fiercest and oldest predators ruling the oceans. One of their greatest features is the formidable speed at which they can flow through the water. Several adaptations have allowed the Shark to swim at such speeds. Patterned surface roughness has been studied and known to alter fluid flow around the body, and Sharks have evolved to develop such a surface which alter the flow in such a way that it reduces the overall drag.

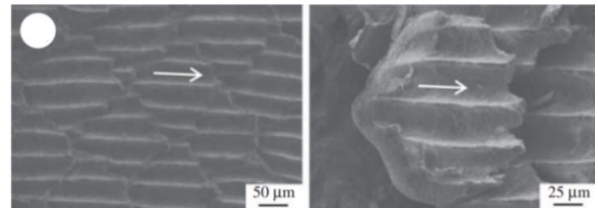

**Figure (1):** Riblet's on a Mako Shark (Martin & Bhushan, 2016) [9]

Figure 1 above shows a microscopic view of the skin of a Mako Shark (Isurus Oxyrinchus). Mako Sharks are a subspecies of sharks which are specifically known for their fast speeds, with a body mass of at least 545 kg, these sharks have recorded speeds up to 74 kmph [1]. Evolution is, in a manner, an optimization that has occurred over millions of years. Similar surface patterns are seen in beaks of skimmer birds, making them more aerodynamic in nature. Biomimicry is the process of imitating biological mechanisms in artificial mechanisms. Several different areas of engineering have adopted biomimetics to improve designs. One of the most prime examples in aerodynamic lies in the development of flight itself. The B2 Stealth Bomber developed by Northrop Grumman was heavily inspired by the shape of the Peregrine Falcon. The development of the delta shaped wing is another example of biomimicry which is inspired from the delta-shaped formation taken by birds when they fly in groups.

Frictional drag (also known as skin friction drag), is drag caused by the friction of fluid against the surface of an object that is moving through it. This drags force delays the forward movement of an aircraft, which leads to more fuel consumption.

The shape and surface of an aircraft are two primary parameters which determine the effect of friction drag on the aircraft. Various changes have been made in the shape and surface of the aircraft. Aircrafts used today are regularly polished and painted to ensure the surface is relatively smooth, similarly, the choice of the airfoil and the use of flaps and slats are optimized to improve the aerodynamic performance. Another type of aerodynamic drag is *form drag* [7], which is caused by the difference in pressure in front of and behind the wing. Boundary layer separation is known to increase the form drag of any object moving through a fluid. Several aerodynamic devices have also been developed such as vortex generators, which induce vortices in the boundary layer to increase its momentum such that the boundary layer separation is delayed [1].

Riblets have always been an area of interest when it comes to reducing aerodynamic drag over an aircraft. A major characteristic of riblets is that they produce secondary flows. Secondary flows are sub-categorized into two types:
1. Prandtl's First Kind
2. Prandtl's Second Kind

The first kind is known as 'skew induced' secondary flow whereas the latter is a product of both skew and stress. Prandtl's second kind can only be induced under turbulent conditions. For riblets, it is difficult to highlight the type of secondary flow being produced and is considered to induce secondary flows of a *third kind* [5], which is driven largely by viscous forces. Several experimental studies Walsh (1983)[15], Lee & Lee (2001)[8] as well as numerical and simulation-based studies Goldstein, 1998; Choi et al, 1993 show that the introduction of riblets concentrates the wall shear stress onto the tip of the riblets therefore decreasing the total area over which the stress acts. From the experimental study on flat plates by Lee & Lee (2001), it was also noticed that the turbulent kinetic velocity and rms velocity fluctuations are greater on ribletted surfaces. Larger velocity and kinetic energy cause the boundary layer separation to be delayed as well. Therefore, it is seen that the implementation of riblets have an effect on both form drag and friction drag on an object.

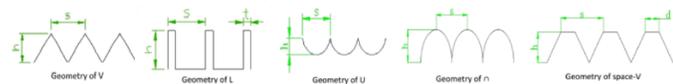

**Figure (2):** Schematic Diagrams of different riblet patterns seen on Sharks (Fu et al., 2017)[4]

Sharks possess several different types of riblet patterns. Some of them have been experimentally and numerically studied to create a comparison. A bio-tribological study conducted by Fu et al. (2017) explored the different riblet shapes as shown in figure 2. The V-shaped riblets are considered the most conventional. There are two main geometric parameters of V-shaped riblets as shown in the figure below. Varying the values of *s* and *h* have been seen to have different effectiveness. From the study of convergent-divergent riblets on a flat plate by Guo et al. (2020), it is seen that by increasing the height of the riblet, the secondary flow becomes much stronger. However, employing larger heights atop airfoils might detriment its function. Another characteristic of the ribletted surfaces is that they induce a certain amount of drag, a study conducted by Raayai-Ardakani and McKinley (2019) highlighted the detrimental effects of non-optimized riblet geometries and concluded that the flow must be of a certain critical Reynolds number for the riblets to act as a drag-reducing component. Ribletted surfaces have been tested in several scenarios, mainly piping technologies and aerodynamic optimization of flat surfaces. However, given the effects of riblets, they can also be implemented on airfoils to delay boundary layer separation. Findings by Zhang et al. (2020) suggest that ribletted surfaces considerably reduce the Reynolds stresses on the upper surface of the airfoil, however it is seen that at sweep angles of 15° the pressure drag increases.

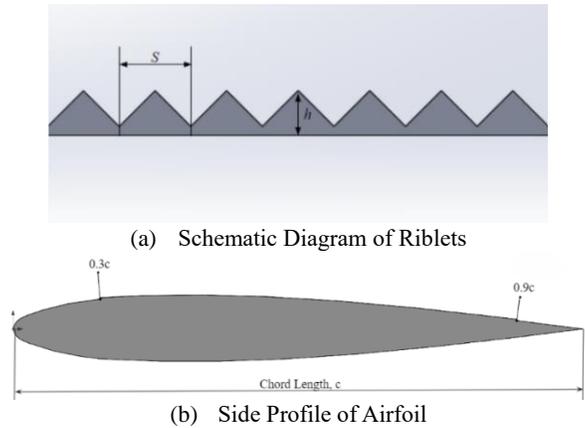

(a) Schematic Diagram of Riblets

(b) Side Profile of Airfoil

**Figure (3):** Schematic Diagrams of different riblet patterns seen on Sharks (Fu et al., 2017)[4]

By using methods of machine learning, optimum geometric parameters can be produced. A form of optimization has been developed based on evolution and the concept of 'survival of the fittest', it is known as a Genetic Algorithm. Genetic algorithms have become very prominent in optimization problems. Genetic algorithms, as seen in biological structures, work using strings of data, where the search is initiated in a pool of points. There are three main operations conducted in genetic algorithms [12]:
1. Reproduction: A specified fitness function determines the success criteria of a string, and a larger fitness value of a string results in a larger probability for it to reproduce.
2. Crossover: This is a process by which a set of strings with high fitness values mate and produce the next generation of points.
3. Mutation: This process is a randomized alteration of any string in the data pool. The main aim of this process is to guard any important characteristic which can get lost in the prior two processes.

Genetic algorithms come under a large umbrella of optimization models known as Evolutionary optimization. Invasive Weed Optimization (IWO) model, Memetic Algorithm and Particle Swarm Optimization are examples of Evolutionary optimization algorithms. These models can be used in conjunction with geometric studies in order to find the optimal shape. A study done by Lulekar et al. (2021) optimizes a valleyed-riblet geometry using a combination of Particle Swarm

Optimization (PSO) and Adaptive Model Refinement (AMR) to obtain drag reductions up to 17.2%.

Using evolutionary algorithms for aerodynamic optimization can open new feats of aerodynamic performance. This paper aims to use an evolutionary algorithm known as the Invasive generation consisted of creating an enclosure from which the airfoil is "snapped" out from. Snappy Hex-Mesh consists of features such as layering controls and zone refinements.

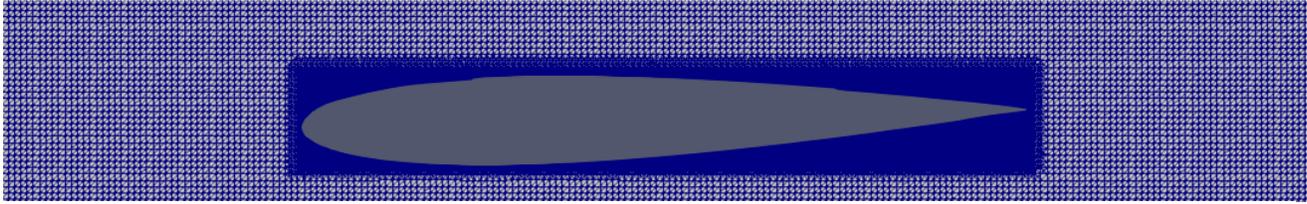

**Figure (4):** Meshing around Airfoil

Weed Optimization (IWO) model to optimize the parameters of these bio-inspired patterns. The invasive weed model is known to produce excellent output by converging close to the global optima. The invasive weed optimization model employs the same mechanism as natural weed colonies. Weeds are plants with an aggressive habit to invade cultivated plants to improve their own growth. The highly adaptive nature of these plants is used as an optimization framework. Similar to the basic genetic algorithm, the IWO model reproduces by making seeds and growing their population until the fittest survives. Comparing to other evolutionary models, it has been seen that IWO works well in escaping local optima as well as the capability to handle complex functions (Mehrabian & Lucas, 2006). In order to pair the IWO with a computational framework, OpenFOAM, which is an open-source Linux-based CFD library, is employed.

This study is presented in three main sections. The methodology is presented where the Computation Fluid Dynamic (CFD) framework, the optimization methods and the geometry generation are discussed in detail. The finding of the simulations is then presented and the various output parameters are discussed. Lastly, the overall achievements and the results from the study are summarized and concluded.

## 2. METHODOLOGY

### 2.1 CFD Framework

The first part of creating a computational framework, the dimensions and the basic geometry is decided. In this study, a NACA 0012 airfoil is chosen as the base airfoil. Using AutoCAD, a basic CAD model is developed where riblets are placed on the upper surface of the airfoil through Boolean commands. The initial CAD is prepared with a chord length of 1 m and 5 riblets of 2mm height. To reduce computational power and decrease the time required for meshing, the base CAD is scaled down such that the chord length reduces to 1 cm. Meshing of the geometry is a key aspect of running a computational simulation. Meshing of the riblet geometry was exported using Snappy Hex-Mesh and Block Mesh through an open-source software known as OpenFOAM. The process for the mesh

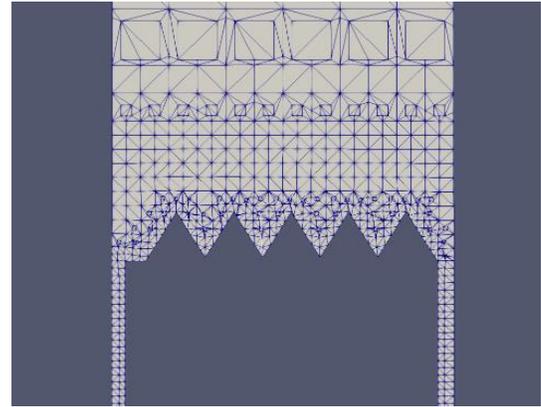

**Figure (5):** Meshing around Riblets

Figures 5 and 6 show the meshing around the airfoil and around the riblets respectively. Zone refinement was added near the airfoil in order to capture the effect of riblets. As evident from the figures above, the mesh created resembles an unstructured mesh. Although for simulations regarding airfoil are conventionally done using a structured C-topology based mesh, in order to mesh effectively around the riblets in 3D, unstructured meshing is used. The mesh consists of a total of 2527511 cells most of which are hexahedral cells (2442387 cells) and the cells forming around the riblets are polyhedral cells (85124 cells). Due to the scaling of the geometry, a small number of faces are flagged with a slightly high skewness of 5.369, however this has minimal effect of the overall simulation.

### 2.2 Solver Settings

Flow simulation has several different models which can be used to predict the flow, and each model is preferred depending on the flow properties. For flow simulation, the Reynolds Averaged Navier-Stokes (RANS) equations are modelled in order to approximate a solution.

$$\frac{\partial(\rho U)}{\partial t} + \nabla \cdot (\rho U U) = -\nabla p + \nabla \cdot [\mu(\nabla U + (\nabla U)^T] + \rho g - \nabla\left(\frac{2}{3}\mu(\nabla \cdot U)\right) - \nabla \cdot (\overline{\rho U'U'}) \quad (1)$$

Equation 1 above signifies the RANS equation, where the last term is known as the 'Reynolds-Stress'. The Reynolds stress needs to be calculated in order to find solution to equation 1. The Reynolds Stress term can be found using the Boussinesq hypothesis given by:

$$-\overline{\rho U'U'} = \mu_t(\nabla U + (\nabla U)^T) - \frac{2}{3}\rho k \boldsymbol{I} - \frac{2}{3}(\nabla \cdot U)\boldsymbol{I} \quad (2)$$

The Boussinesq hypothesis links the Reynolds stress to the mean velocity gradient with a coefficient $\mu_t$, which is called Eddy Viscosity. To approximate the value of the Eddy Viscosity, several models are present. Since the simulation requires extensive focus near the surface of geometry, the k-ω model is used as a solver. The k-Ω solver is more accurate towards the near wall boundary layers as compared to the k- ε turbulence model. Both these models are two-equation models used to approximate the *eddy viscosity*. The k-ω turbulence model has evolved over decades with better forms of approximations. The model used in this study is the Shear Stress Transport (SST) formulation. The SST model has the added advantage of working well under adverse pressure gradients as compared to its more primitive 'Wilcox's k-ω model'. The SST model has itself evolved through the years, the version of OpenFOAM utilized for this study uses the Menter SST model from 2003. The model describes the eddy viscosity as:

$$\mu_t = \frac{\rho a_1 k}{max(a_1\omega, SF_2)} \quad (3)$$

The transport equations for the k-ω model, defines the kinetic turbulent energy, k, and the specific dissipation rate, ω. Below are the modelling equations for k and ω respectively:

$$\frac{D}{Dt}(\rho k) = \nabla \cdot (\rho D_k \nabla k) + \rho G - \frac{2}{3}\rho k(\nabla \cdot u) - \rho\beta^*\omega k + S_k \quad (4)$$

$$\frac{D}{Dt}(\rho\omega) = \nabla \cdot (\rho D_w \nabla \omega) + \frac{\rho\gamma G}{v} - \frac{2}{3}\rho\gamma\omega(\nabla \cdot u) - \rho\beta\omega^2 - \rho(F_1 - 1)CD_{k\omega} + S_\omega \quad (5)$$

To initialize the model, the turbulent kinetic energy and the specific dissipation values are included in the initial case. The turbulence is considered isotropic and is calculated using:

$$k = \frac{3}{2}(I|u_{ref}|)^2 \quad (6)$$

Accordingly, the specific dissipation can be calculated by:

$$\omega = \frac{k^{0.5}}{C_\mu^{0.25} L} \quad (7)$$

Here $C_\mu$ is a constant with a magnitude of 0.09. After these values are calculated the case can proceed. However, to solve these equations, certain solving algorithms have been developed. The algorithm used in for this study is the SIMPLE model, which is a pressure based solver which stands for "Semi-Implicit Method for Pressure Lined Equations" [2]. This method is a considered quite efficient due to its stability and uses relatively lower computational effort, especially to model simpler flows, a faster convergence can be obtained.

**2.3 Invasive Weed Optimization**

The invasive weed optimization algorithm (IWO) is a population-based evolutionary optimization algorithm that finds the general optimum of a mathematical function through imitating compatibility and randomness of weeds colony. This method is inspired by a phenomenon in agriculture called colonies of invasive weeds. The IWO tries to simulate the colonizing behaviour of invasive weeds in nature, by utilizing some of their fundamental properties which leads to a robust optimization algorithm. Weed is a plant that grows unintentionally. They invade the cropping system (field) by utilizing the unused resources from the opportunity spaces created in the fields. Weeds might have many uses and benefits in some regions, if the same plant grows in a region that interferes with human needs and activities, it is called a weed. This methods works as:

*Primary Population Initialization*: A limited number of seeds are distributed in the search space and their fitness value is calculated. They are ranked on the basis of fitness value.

*Reproduction*: Each seed grows into a flowering plant and produces seeds that depend on their fitness value. The number of grains of grasses keeps on decreasing linearly. Further new seeds are dispersed over the design solution space.

*Spatial Dispersion*: The seeds produced by the group in the normal distribution with a mean planting position and standard deviation (SD) are produced with the help of an equation. The conversion assures that the fall of a grain in the range decreases nonlinearly at each step, leading to more fit plants and eliminating inappropriate plants.

*Competitive Deprivation*: If the numbers of grasses exceed the maximum numbers of grasses in the colony, the grass with worst fitness is removed from the colony so that a constant numbers of herbs are remained in the colony.
This process continues until the maximum number of iterations is reached, then the minimum colony cost function of the grasses is stored. If it has not reached the maximum number of iterations then the process is restarted from the point where they are reproduced based on individual's fitness value.

**2.3 Geometry Generation**

To set up the computational fluid dynamic framework, a basic CAD model was generated using the Fusion 360 suite. NACA-0012 airfoil plots are procured and fed into the CAD suite which is then extruded to an appropriate depth. This extrusion depth is maintained at 0.015mm, initially. The 3-D model has been scaled such that the chord of the airfoil is of unit length. Thereafter, we have sketched our triangular-shaped riblets in a plane normal to that of the airfoil. We have then extruded the triangular profile along the path indicated by the upper portion of the airfoil sketch with the help of the sweep tool. With the help of the 'Pattern on Path' tool, 4 more extruded profiles have been made over the top surface of the airfoil. Finally, a NACA Airfoil with 5 riblets over it is created. This CAD is used as an initial test to mesh generation and see if the computational fluid dynamic analysis reflects the required features. Furthermore, until the auto-manipulable CAD is readied, the initial CAD is used to develop the CFD case and to improve its characteristics such as the mesh volume, and mesh quality, to reduce the meshing time whilst keeping its quality high.

A Python-based code is written in which the height and wavelength of riblets are initialized by the user and it will then automatically generate its 'STL' file that will have riblets over the Airfoil. The process for creating this python code is as follows: firstly, the CAD file of the airfoil which was made with the help of Fusion 360 earlier is converted into an STL format file. Python libraries such as Solid python, NumPy, os, and stl, allow this initial stl file to be read and then manipulated. The endpoints of the riblets are fixed as per the parameters of the airfoil. The code takes the riblet height and wavelength as input from the user and then generates triangle coordinates between the endpoints of riblets. These triangular profiles are then extruded along with the coordinates of the top surface of NACA Airfoil which are provided in the code. This extrusion prepares a 'scad' file which is then converted into an STL file. Figure 7 shows the generated STL file of the riblets. This STL file is then combined with the STL file of NACA Airfoil using the codes available in the python library.

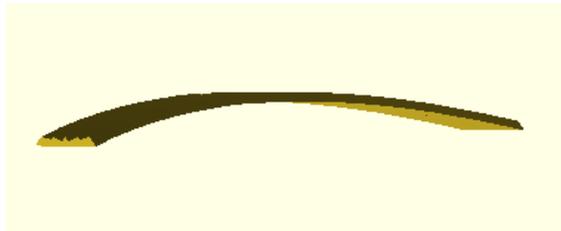

**Figure (6):** OpenSCAD Generated Riblets STL

### 2.4 Overall Optimization Framework

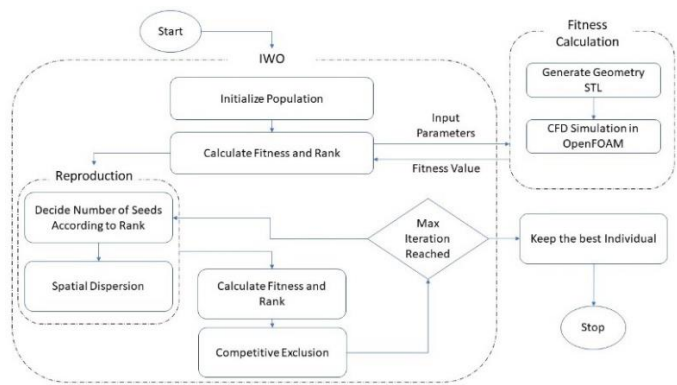

**Figure (7):** Optimization Flowchart

## 3. RESULTS AND DISCUSSION

### 3.1 Grid Dependency Test

In order to assess the reliability of the mesh used for the simulations, a grid dependency test is performed. Since the flow alterations occur near the riblets, the velocity field at 0.4c is tested in a high Reynolds (Re) flow field. A well-established mesh setting is essential for the study both in order to accept the changes in the riblet parameters as well as to capture the flow field near the riblets. Figure 5 below shows the changes in velocity along the height of the riblet.

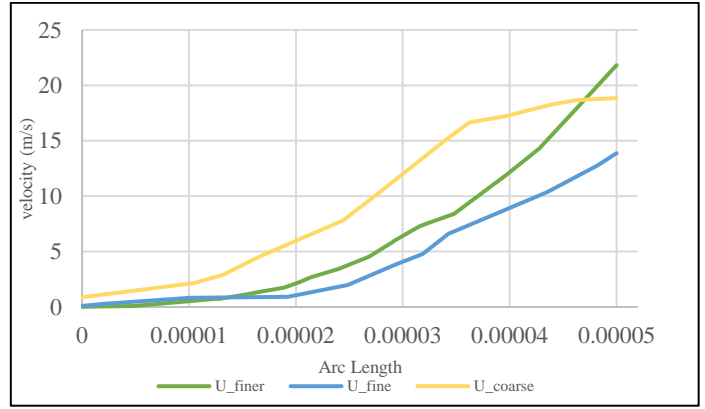

**Figure (8):** Grid Dependency Test

It is evident that a coarse mesh overshoots the values and fails to capture the velocity field as well as the finer two meshes. The fine mesh is able to capture the velocity trend and the velocity values fall close to the values generated from the finer mesh.

The IWO optimization is done over 102 iterations. The figure below depicts the trend of the Coefficient of Drag ($C_D$) over these iterations.

### 3.2 IWO Results

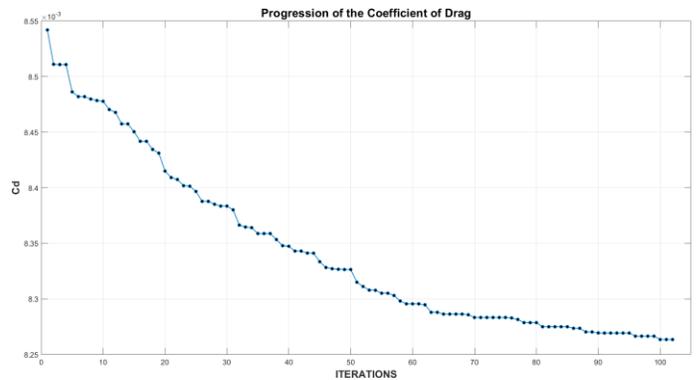

**Figure (9):** $C_D$ evolution over Iterations

The graph above shows how the $C_D$ of the airfoil decreases with every iteration. The algorithm, iterates over several generations and picks the geometry which gives the lowest $C_D$ value. The initial geometric parameters were set as 2.0 mm riblet height and 2.5 mm wavelength. The optimization occurs over 102 iterations with an overall decrease in $C_D$ of 3.26%. The optimized riblet parameters are simulated singularly.

### 3.3 Optimized Riblet Geometry

It is seen that the optimized riblet geometry has a riblet height of 0.77mm and a wavelength of 1.24mm. The generated geometry is then simulated under the same conditions providing a $C_D$ of 0.00826. Comparing it to the $C_D$ achieved from a bare airfoil, 0.00909, a decrease of 7.59% is observed. This decrease in $C_D$ can be further explained with the help of the contour plots.

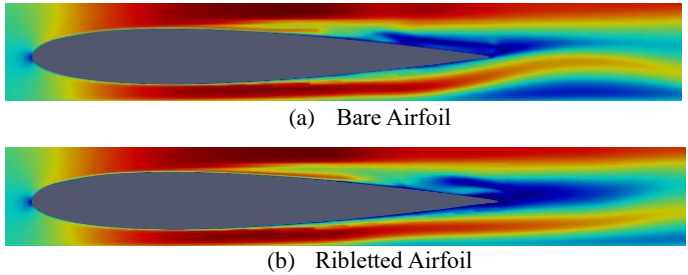

(a)  Bare Airfoil

(b)  Ribletted Airfoil

**Figure (10):** Velocity Contours (a) Bare Airfoil; (b) Ribletted Airfoil

Figure 8 and 9 represent the velocity fields over a bare airfoil and a ribletted airfoil respectively. It is evident that the boundary layer separation is delayed in the ribletted airfoil as compared to the bare airfoil. Secondly, a stronger boundary layer reattachment zone is created in the ribletted variant. Boundary layer separation is one of the major factors that increase the overall drag of an airfoil due to the formulation of pressure drag. Riblets are able to induce small scale secondary flows on the upper surface which increases the overall momentum and kinetic energy within the airfoil's boundary layer. Having a larger momentum and kinetic energy reduces the tendency of the boundary layer to separate. A delayed separation reduces the wake generated behind the airfoil, which creates a highly turbulent and low pressure region.

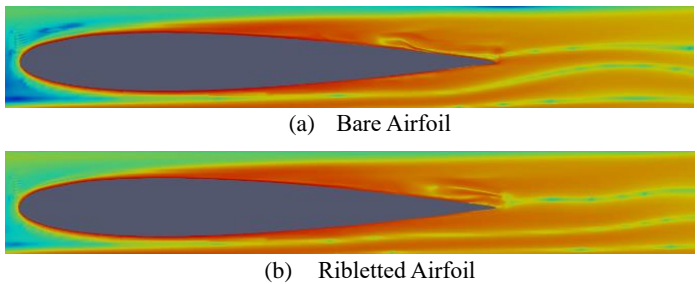

(a)  Bare Airfoil

(b)  Ribletted Airfoil

**Figure (11):** Vorticity Contours (a) Bare Airfoil; (b) Ribletted Airfoil

Figures 10 and 11 can be used to compare the vorticity magnitudes on the bare airfoil and the ribletted airfoil. The vorticity vector can be mathematically represented as:

$$\Omega = \nabla \cdot u \qquad (8)$$

Using vorticity, the levels of turbulence and flow behavior around the airfoil can be studied. From the simulations, it was seen that the bare airfoil contained stronger vorticities in the wake region of the airfoil. From the simulations, it was also seen that the riblets generated 18.9% stronger vortices in the boundary layer region of the airfoil.

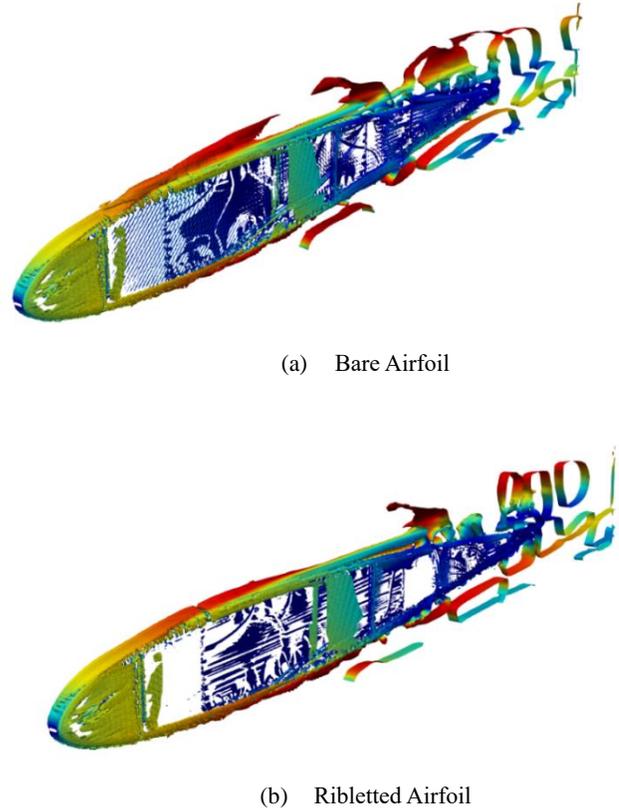

(a)  Bare Airfoil

(b)  Ribletted Airfoil

**Figure (12):** Q-Criterion (a) Bare Airfoil; (b) Ribletted Airfoil

Q-criterion is another plot which can be used to magnify the findings from the simulation. Figures 12 and 13 above show the Q-criterion values of bare and ribletted airfoils respectively. It is evident that the ribletted airfoil is able to keep the flow attached to the surface for a longer distance than the bare airfoil.

Three specimen were further analyzed by comparing their $C_D$ against the angle of attack (AoA). The bare airfoil, the initial riblet parameters and the optimized parameters are simulated under different angle of attacks ranging from -2º to 6º.

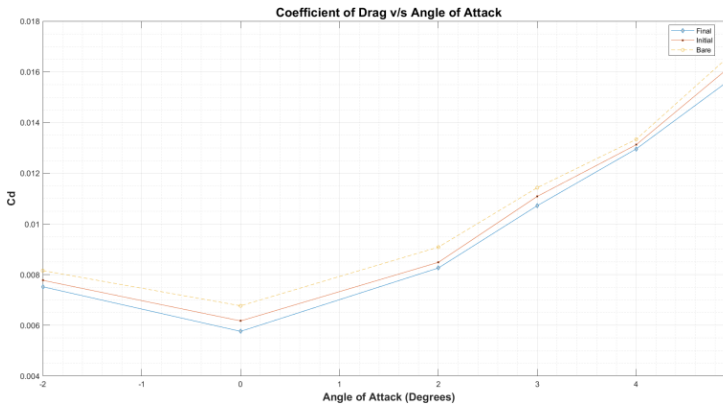

**Figure (13):** $C_D$ vs AoA

It is visible that all three airfoils follow a similar trend in each angle of attack, however, the addition of riblets decreases the amount of drag over the airfoil. A substantial decrease in the $C_D$ is noticed with the addition of the optimized riblets.

## 4. CONCLUSION

In this paper the utilization of saw-tooth riblets are studied and it's geometric parameters are manipulated in order to achieve maximum drag reduction. As the theory follows, the riblets induce a secondary flow over the airfoil which helps delaying the boundary layer separation by increasing the energy and momentum of the flow in the boundary layer. It is seen from several studies that the shape and the geometric parameters have a significant effect on the effectiveness of the riblet. The use of a metaheuristic optimizer is considered for the optimization of these geometric parameters. It is seen that the use of Invasive Weed Optimization (IWO) is a viable method of optimizing the geometric parameters of saw-tooth riblets. Riblets being an bio-mimetic device allows a considerable drag reduction on the NACA-0012 airfoil. The IWO optimization conducted over 102 iterations is able to provide a riblet geometry with an h+/s+ value of approximately 0.48 with sufficient drag reduction. Furthermore, with the use of the optimized riblets an overall drag reduction of 7.59% was achieved.

**ACKNOWLEDGEMENTS**

The authors like to thank the members of the Fluid Mechanics Group & staff of Fluid Mechanics Lab at Delhi Technological University for providing required resources and help in troubleshooting the problems faced while running numerical Simulations.